\documentclass[sigconf]{acmart}
\usepackage{gensymb}
\usepackage{xspace}
\usepackage{manfnt}
\usepackage{hyperref}
\hypersetup{
    colorlinks=true,
    linkcolor=blue,
    filecolor=magenta,      
    urlcolor=cyan,
    pdftitle={Computing within LIMITS},
    pdfpagemode=FullScreen,
    }
\usepackage[vskip=.7em,font=itshape,leftmargin=2em,rightmargin=2em]{quoting}
\usepackage{eurosym}
\usepackage{relsize}

\newcommand{\mondegree}{\degree{C}\xspace}
\newcommand{\ts}{\textsuperscript}
\newcommand{\coo}{CO\textsubscript{2}\xspace}
\newcommand{\cooeq}{\coo{}eq\xspace}
\newcommand{\mytranslation}{translated by the authors\xspace}
\newcommand{\mybold}{bold typeface from the authors\xspace}
\newcommand{\myboldandtranslation}{translation and bold typeface from the authors}
\newcommand{\life}{LIFE\xspace}
\newcommand{\ssdc}{small-scale data center\xspace}
\newcommand{\SSDC}{Small-Scale Data Center\xspace}
\newcommand{\etal}{\emph{et al.}\xspace}

\AtBeginDocument{%
  }

\setcopyright{none}
\acmDOI{}
\acmISBN{}  
\acmConference[LIMITS'25]{11th Workshop on Computing within Limits}{June 26--27, 2025}{Online}

\makeatletter
\def\hang{\hangindent\parindent}
\def\d@nger{\medbreak\begingroup\clubpenalty=10000
  \def\par{\endgraf\endgroup\medbreak} \noindent\hang\hangafter=-2
  \hbox to0pt{\hskip-\hangindent\dbend\hfill}}
\outer\def\danger{\d@nger}
\makeatother

\begin{document}

\title{ICT Within Limits Is Bound To Be Old-Fashioned By Design}

\author{Olivier Michel}
\email{olivier.michel@u-pec.fr}
\affiliation{%
  \institution{LACL - Université Paris-Est Créteil}
  \city{Créteil}
  \country{France}}

\orcid{0009-0005-4468-0161}

\author{Émilie Frenkiel}
\email{emilie.frenkiel@u-pec.fr}
\affiliation{%
  \institution{LIPHA - Université Paris-Est Créteil}
  \city{Créteil}
  \country{France}
}

\renewcommand{\shortauthors}{Michel, O. \& Frenkiel, É.}

\begin{abstract}


Crossing multiple planetary boundaries places us in a zone of uncertainty that is characterized by
considerable fluctuations in climatic events.  The situation is exacerbated by the relentless use of
resources and energy required to develop digital infrastructures that have become pervasive and
ubiquitous. We are bound to these infrastructures, dead technologies and negative commons, just as
much as they bind us. Although their growth threatens the necessary reduction of our impact, we have
a responsibility to maintain them until we can do without them.

In university setting, as well as in any public organization, urban mines \emph{per se}, we propose
an IT architecture based on the exclusive use of unreliable \emph{waste from electrical and
electronic equipment} (WEEE) as a frugal alternative to the unabating replacement of
devices. Powered by renewable energy, autonomous, robust, adaptable, and built on battle-tested
open-source software, we envision this solution for a situation where use is bound to decline
eventually, to close this damaging technological chapter. Digital technology, the idol of modern
times, is to meet its twilight if we do not want to irrevocably alter the critical zone.

\end{abstract}




\maketitle

\section{A Major Environmental Crisis} \label{sec:uncertainty}

As the IPCC states~\cite{ipcc_climate_2021}, ``climate change is widespread, rapid and
intensifying''. We are facing a significant increase in number and intensity of climatic events like
extreme temperatures, drought conditions, heatwaves, fires, flooding, glacier and permafrost
melting... Relative to the pre-industrial era, we face an average global temperature increase of
+1.15\mondegree, with local values highly dependent on the regions of the world considerd; for
example, France faces a higher value with an increase of +1.7\mondegree due to its geographical
characteristics. As is the case on a regular basis, each passing month brings record-breaking
temperatures.\footnote{For example, in 2024, April that was the 11\ts{th} consecutive warmest month
globally~\cite{copernicus_april_2024}.} 2024 was the first calendar year with a global average
temperature exceeding 1.5\mondegree above pre-industrial level~\cite{copernicus_global_2025}.


Unfortunately, as we know, climate change isn't the whole story. The planetary boundaries'
framework~\cite{rockstrom_safe_2009}, which
\begin{quoting}{}
delineates the biophysical and biochemical systems and processes known to regulate the state of the
planet [...] to maintain Earth system stability and life-support systems conducive to the human
welfare [...]
\end{quoting}%
reminds us that life on earth relies on additional properties. Interactions between the geosphere
and the biosphere have controlled environmental conditions for over 3 billions years, with the
Holocene that started about 11,700 years ago being rather stable.

Climate change is one of the 9 planetary boundaries\footnote{Six of which have already been crossed,
with a 7\ts{th}, ocean acidification, being currently crossed.} which includes Greenhouse Gases
(GHG) expressed in terms of \cooeq concentration and radiative forcing. If we consider \coo, the
current value is around 422.5 ppm~\cite{noauthor_trends_2025}
and~\cite{friedlingstein_global_2024}[p. 7], way above the 350 ppm boundary and the 278 ppm value
of the pre-industrial era.
In 2021, IPCC modelization in its sixth assessment report~\cite{lee_ipcc_2023}[pp. 20-21] states
that (\mybold)
\begin{quoting}{}
For every 1,000 Gt\coo emitted by human activity, global surface temperature rises by 0.45\mondegree
[...] remaining carbon budgets from the beginning of 2020 are 500 Gt\coo [...] \textbf{The stronger
the reductions in non-\coo emissions}, the lower the resulting temperatures are for a given
remaining carbon budget [...]
\end{quoting}

Four years later, in 2025, June the 19\textsuperscript{th}, at the time of writing, the remaining
\coo budget for a 50\% likelihood to limit global warming to 1.5\mondegree above the 1850-1900 level
(Paris' agreement and SSP1-1.9) has been reduced to merely 130
Gt~\cite{forster_indicators_2025}[Table 8 p. 2663]. This budget will be exhausted in 3 years if
global \coo emissions remain at 2024 levels (about 42 Gt\coo\ yr$^{-1}$).


Multiple crossed planetary boundaries and, among them, low \coo budget, places us in a \emph{zone of
uncertainty} that is characterized by \emph{considerable fluctuations} in climatic events. Moreover,
IPCC projections commit us to \emph{reducing our net GHG emissions}, or at least do the best we can
to \emph{avoid making the situation worse}.

\section{ICT, an Immaterial Limitless Technology} \label{sec:ict-no-limit}

\subsection{Exponential Laws with No Counterparts} \label{sec:laws}

Various \emph{laws} in ICT, namely 
\begin{itemize}
 \item Moore's law~\cite{moore_cramming_2006} on the growth of the density of transistors
       integration into CPUs together with
 \item Dennard's scaling law~\cite{dennard_design_2007} on the power density of a circuit - with
       transistors scaling down - remaining constant (at constant surface area)\footnote{The authors
       would like to thank G. Roussilhe for bringing this point to their attention,
       see~\cite{roussilhe_phase_2025}.},
 \item Kryder's law~\cite{walter_kryders_2005} on disk storage capacity, 
 \item Metcalfe's law on the network effect~\cite{metcalfe_metcalfes_2013},
 \item Nielsen's law on the state of the art of available bandwidth~\cite{nielsen_nielsens_nodate},
\end{itemize}
combined with software engineering principles based on
extensibility~\cite{maraninchi_planetary_2024}, have built a disciplinary field structured on the
exponential availability of resources with no physical
counterpart\footnote{In~\cite{chapoutot_grand_2021}, the historian J. Chapoutot coined the neologism
``illimitisme'' -- which can be translated as \emph{limitlessness} -- to describe precisely the
pursuit of endless technological progress through the unlimited exploitation of available
resources. The term recalls the notion of ``illimitisme patriarcal'' developed by F. d'Eaubonne,
who, as early as 1974 in her work on the construction of the concept of
ecofeminism~\cite{eaubonne_feminisme_2020}, already pointed to the exploitation of available
resources.}. An economist's dream in action!  This very state of mind is reflected in the name
coined for the generalization of grid computing, \emph{cloud computing}\footnote{Ironically enough,
if you think about it for a moment, the name itself was a harbinger of ICT's major climate effects,
as clouds are made of water droplets condensing from water vapor, which is the main (natural) GHG
accounting for 15\% of the Earth's GHG effect.} where resources are supposedly in an ether,
unlimited on-demand (hyperscale), without any constraint.

\subsection{The Unavoidable Reality Principle} 

\begin{figure}[!t]
  \centering
  \includegraphics[width=.7\linewidth]{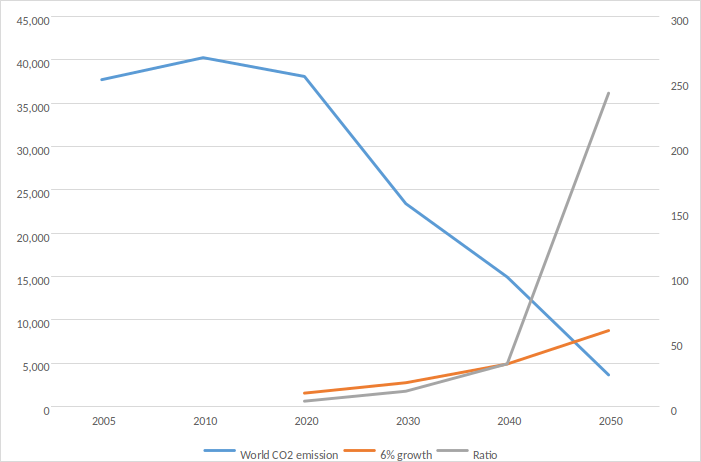}
  \caption{IPCC scenario SSP1-1.9 and \emph{lower bound of ICT growth} (6\%). Left ordinate are MT of \coo
  emission per year; right ordinate are percentages for the grey curve.}
 \label{fig:growth6}
\end{figure}

But the phantasm of the absence of materiality eventually collides with the materiality of the
infrastructure and devices that enable the use of digital (non-convivial)\footnote{This brief
analysis of the (non)materiality of digital tools should be analyzed more thoroughly from the
perspective of conviviality proposed by I. Illich~\cite{avant_tools_1975} and from
A. Gorz~\cite{gorz_immaterial_2010}.} tools: the most conservative
estimates~\cite{the_shift_project_pour_2018} put the number for ICT devices to 34 billions with 4.1
billions users requiring about 5.5\% of the world's electricity. All those devices communicate using
submarine cables, among others, for a total of about 1.48 million km (as of 2025) with length
ranging from 131 km (from Ireland to UK) to 20,000 km (from Asia to
America)~\cite{telegography_submarine_2025}; smartphones require more than 50\% of periodic table of
the elements~\cite{systext_metaux_2017} with rare-earth elements production having substantial
geopolitical and environmental impact~\cite{ganguli_rare_2018,zapp_environmental_2022}.

\subsection{Recycling Won't Help} \label{sec:recyclage}

Many growth scenarios for ICT are based on significant ability to recycle essential elements found
in devices. Recycling takes place at the final step of a product's life cycle (which consists in the
5 following phases: raw material extraction, manufacturing and processing, transportation, usage and
retail, waste disposal). Unfortunately, 100\% recycling is
impossible~\cite{systext_controverses_2024}[p. 4] as ''[...] the life cycle of metal is most often a
succession of material and energy losses, subject to physical and thermodynamic limits.''
(\mytranslation). Many rare-earth elements that are crucial for ICT devices have very low
\emph{end-of-life recycling rates}\footnote{Abbreviated EOL-RR.}, high \emph{loss rate}\footnote{The
\emph{loss rate} (in kg lost per kg of metal extracted per year) represents the rate at which
extracted metal becomes unavailable for further use. It is calculated as the inverse function of the
average service life. (\mytranslation from footnote 207 page 107).} and short \emph{service life of
metal}\footnote{The \emph{service life} of a metal (in years) represents the average duration it is
used in the economy, from the time it is mined until it is completely lost to landfill or the
environment, so that it becomes unavailable for further use. (\mytranslation from footnote 206 page
107).} (see figure 3 and 5
of~\cite{charpentier_poncelet_losses_2022}). In~\cite{systext_controverses_2024}, the authors remind
us of the ``crucial role of \emph{lengthening the lifespans of products} to improve the conservation
of metals in the economy''.

One might wonder what would happen if, for a given metal, both EOL-RR (say 80\%) and service life
(say 10 years) where high and a reasonable (yet exponential) constant annual consumption growth rate
of 4\% would drive the use of that metal? The net effect of recycling would only \emph{shift} the
consumption of that metal for only 20 years.\footnote{See~\cite{systext_controverses_2024}[figure 77
page 164].} In a \emph{growing economy}, recycling is not the appropriate response to the issues
raised by the reduced availability of rare-earth elements used in ICT devices.

%

\subsection{Incompatibility of ICT Growth of GHG Emissions with The Paris Agreement} \label{sec:ictgrowth}

\begin{figure}[!t]
  \centering
  \includegraphics[width=.7\linewidth]{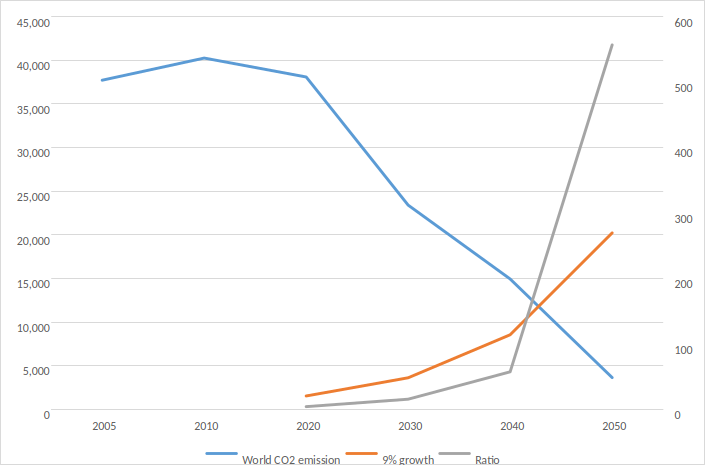}
  \caption{IPCC scenario SSP1-1.9 and \emph{upper bound of ICT growth} (9\%). Left ordinate are MT of \coo
  emission per year; right ordinate are percentages for the grey curve.}
 \label{fig:growth9}
\end{figure}

Based on~\cite{the_shift_project_lean_2018,freitag_real_2021}, ICT is accountable for 5-6\% of world
primary energy consumption and between 2.1\% and 3.9\% of GHG emissions, \emph{with 30\% coming from
embodied emissions}. Based on data for 2015-2019, the annual growth of these figures are between 6\%
(lower bound) and 9\% (upper bound), that is, before the availability of generative IA.

\begin{figure}[!h]
  \centering
  \includegraphics[width=.7\linewidth]{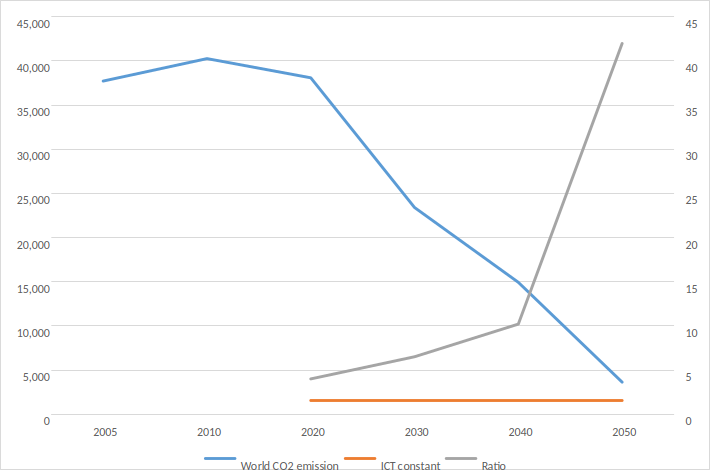}
  \caption{IPCC scenario SSP1-1.9 and \emph{constant ICT}. Left ordinate are MT of \coo
  emission per year; right ordinate are percentages for the grey curve.}
 \label{fig:constant}
\end{figure}

Following the analysis suggested by D. Trystram, Y. Malot and G. Raffin at Université Grenoble
Alpes\footnote{Their app is available at
\url{https://edge-intelligence.imag.fr/trajectory_app.html}.}, figure~\ref{fig:growth6}
(resp.~\ref{fig:growth9} and~\ref{fig:constant}) simulates \coo world and ICT levels in IPCC's
scenario SSP1-1.9 with a minimal (resp. maximal and constant) ICT growth of 6\% (resp. 9\% and 0\%)
per year. The blue curve is the world \coo emission in Mt/year as it should be to comply with the
Paris agreement; the red curve is ICT's share of \coo emissions; the grey curve is the ratio
between the blue and the red curve in percentage (the blue curve includes the red curve in 2010, but
they are independent for the rest of the simulation). It is clear that lower and upper bounds of ICT
growth are absolutely not compatible with the Paris agreement. Even keeping ICT constant to its
current share of GHG emissions does not enable compliance with the Paris agreement.

Clearly, the share of ICT in global GHG emissions must be reduced~\cite{pargman_moores_2020} in the
same way as other industries\footnote{We won't go into the controversy over \emph{decoupling}, which
would enable growth while reducing environmental impact, but would not address the problems of raw
material depletion mentioned in section~\ref{sec:recyclage}; interested readers can refer to the
work of T. Parrique~\cite{timothee_parrique_academic_nodate}.}. If limits have to be set to the
expansion of digital technology, and new fields of research explored (these two issues being
discussed in the international workshop~\cite{nardi_computing_2018}
and~\cite{noauthor_undone_nodate}), there is still a need to provide digital services, given their
level of intricacy and pervasiveness in today's societies, but with the lowest possible
environmental impact and without generating (direct or indirect) rebound effects. Considering
digital technologies in its \emph{infantile stage of development}, introducing limits leading to the
gradual reduction of digital services, that is going past the peak
ICT~\cite{tomlinson_collapse_2012}, if it is to be considered desirable, cannot be carried out
abruptly; to paraphrase B. Latour, the landing on earth~\cite{latour_ou_2017} should be gradual,
collectively decided and fair.

\section{Negative Commons, Zombie Technology and Attachments} \label{sec:negcommons}

As we have just seen, we are facing a bleak situation: climate change requires us to reduce GHG
emissions, but it also increases the uncertainty and fluctuations of climatic events. At the same
time, we have inherited a \emph{large technical system}~\cite{ellul_systeme_1977} - digital
technologies - that permeates all levels of social organization. How can we understand this legacy,
and what can we do with it? A few concepts help us to better grasp the particular situation in which
we find ourselves, and how we can navigate through it: \emph{negative commons}, \emph{zombie
technology}, \emph{attachments} and \emph{adaptability}.

\subsection{Negative Commons}

The \emph{commons}\cite{ostrom_governing_1990} consist of a triptych (focused on monopolization): an
incommon fraught with conflict - a resource that we wish to share; a resource managed by a community
(which is neither the state nor a private firm); which sets up rules and governance for this
purpose. This resource has a \emph{utility}, it has \emph{positive effects}. But what about the
\emph{negative effects} produced by certain things? Let's take the paradigmatic example of nuclear
power plants: we \emph{inherit} these infrastructures that we can't live without, they have a
limited lifespan, they're impossible to dismantle\footnote{France's Brennilis plant, shut down in
1985, is a case in point. Its end is constantly being put off, and is scheduled for 2040, at an
estimated cost of around 850 million \texteuro.}. We therefore need to extend their lifespan.  We
could simply say that a nuclear power plant is a waste product, but it's a special kind of waste,
one that cannot be reintegrated into natural bio-geo-chemical cycles. It's not the economists'
\emph{negative externality}: it's not an unintended result, but on the contrary a \emph{condition of
possibility} for cheap energy that makes the construction of digital infrastructures possible. The
extraction of precious metals by children, the war in the DRC, the former Agbogbloshie landfill in
Ghana, the poor protection afforded to the workers who make smartphones, etc. are not side effects
of their production: at this price on these markets, they are quasi-necessities, constitutive
elements and not unfortunate consequences.

A. Monnin \etal propose the notion of \emph{negative
commons}~\cite{bonnet_heritage_2021,monnin_planetary_2021} to rethink the commons in the light of
thoughts on the Anthropocene, and to move away from a solutionist vision in which the commons would
save the world. Indeed, while the issue of the commons concerns the means of avoiding the
appropriation of ''common'' realities, or of reappropriating what has been captured by enclosures,
there remain the realities that nobody wants (organic and nuclear waste, technosphere waste,
abandoned infrastructure, polluted soil, dried-up rivers, etc.). These are the ruins that fall into
two categories: \emph{ruined ruins} (called \emph{ruina ruinata}), which escape any desire for
appropriation (such as picturesque and romantic ruins) and \emph{ruinous ruins} (called \emph{ruina
ruinans}) which are always in action.

The devices that dig them out, the economic models that make them profitable, the supply chains that
export them to the four corners of the planet... these are the \emph{negative commons we inherit}.
These realities (technical, managerial, economic, logistical, etc.) are negative commons that we are
inheriting, because an ever-growing proportion of the world's population is linked to them in the
short term, even though their operation constitutes the greatest threat to the planet's habitability
in the medium term. These ruins are not to be found in an imaginary world of decline or decadence,
but in the gleaming, high-tech realities of the destructive regime of intensive innovation at every
turn, and the incessant renewal it demands. It's the ruin that is still productive: productive of
new ruins, ruinous or ruined. Some ruins are both ruined and ruinous, such as oil and its iconic
wells, which are both ruined because they are the product of hundreds of millions of years of
transformation of organic matter, and ruinous because they are a miracle product for industry
(transforming the world) and a source of massive GHGs emissions.

\subsection{Zombie Technologies}

What characterizes ruins is their persistence over time. Certain technologies, involve finite
resources (energy, metals, etc.) as opposed to the CHNOPS\footnote{Acronym for Carbon, Hydrogen,
Nitrogen, Oxygen, Phosporus and Sulfur.} chemical elements that make up living matter and are
sustainable because they are renewable. These resources are drawn from available stocks, the most
critical of which do not have a satisfactory recycling rate (see section~\ref{sec:recyclage}). These
technologies are doomed to survive in a degraded form for a very long time. This is why
J. Halloy~\cite{halloy_au-a_2020} refers to them as \emph{zombie technologies} (or \emph{dead
technologies}, as opposed to \emph{living technologies}) because they do not want to die or
disappear. Activities can be zombified, such as agriculture, which is 14,000 years old and which, in
less than a century, has been totally transformed (machinery, inputs, drones, etc.). In 2018, 9\% of
German farmers used drones (according to a study conducted by the DBV~\cite{noauthor_9_2018}).

Zombie techniques give the impression of being alive, through their frenetic activity, but are
undermined from within: a death in waiting. J. Halloy proposes three criteria that favor
zombification~\cite{triclot_prendre_2024}[p. 299]: the use of finite stocks that impose a time
limit on the activity; the use of a power exceeding the capacities of the environment in which this
technique is used; the generalization of these characteristics on a large scale (and we should add
that they take part to the current ecocide, following~\cite{comber_computing_2023}). It will be easy
to recognize that all these characteristics are present in ICT. But these technologies pose an
additional problem, that of our \emph{attachments}.

\subsection{Attachments, De-attachments and Re-attachments} \label{sec:attachements}

We are \emph{attached}~\cite{callon_ni_1999,hennion_sociologie_2004,hennion_vous_2010} to the ruins,
which requires us to question our values: \emph{what we care about} and what we are \emph{attached
to}. This twofold movement is important for understanding the importance of ICT and the need to make
it last over time. This attachment to our living conditions and our infrastructures is essential,
even if they are negative and lead us to disaster in their current state (see
section~\ref{sec:ictgrowth}). Because we don't have the choice to make them work differently, we
have the obligation to make do with them (as is the case with the digitization of a large number of
essential public services, for example). This requires working on \emph{de-attachment} and
\emph{re-attachment}, on a collective scale. The challenge is to make the non-political political in
order to maintain habitable ecosystems. We face multiple problems:
\begin{itemize}
 \item the scale (spatial/temporal/functional/organizational/etc.) of negative commons;
 \item the democratic challenge of determining the negative character of certain commons;
 \item not letting communities alone to manage negative commons, not relying on the resilience of
       populations or the administrative management of crises;
 \item technosphere/biosphere opposition: asking the question of their future, which threatens our
       future existence as much as it makes our current existence possible;
 \item the imperative need to maintain and take care of the ruins that are not yet ruined (roads,
       bridges, dams, ..., ICT)
\end{itemize}
While A. Monnin advocates an ecology of dismantling and closure~\cite{monnin_politiser_2023}, we
believe that with regard to ICT, it is important, as we have said several times before, to keep
these zombie infrastructures alive and in working order for a while longer (a very interesting
initiative from~\cite{franquesa_devices_2018}, which is not contradictory to our point of view, is
to consider smartphones as \emph{commons}, implying a form of collective reappropriation, in a
perspective that could enable the development of \emph{rights} associated to
\emph{things}~\cite{denis_soin_2022} -- as opposed to the notion of \emph{objects}) -- to ensure a
safe landing for societies that heavily rely upon them.

\subsection{Technical Systems Focused on Efficiency Gains} \label{sec:hamant}

In the development of \emph{large technical systems}, after the Renaissance and the Enlightment,
modern societies focused on efficiency and growth. This technical choice has had significant
consequences on the type of society that has evolved. Quoting O. Hamant~\cite{hamant_tracts_2024}
(\myboldandtranslation)
\begin{quoting}{}
 [...]  \textbf{if everything is constantly subjected to change}, effectiveness and efficiency are
the instruments of an optimization that locks us into a narrow and therefore inadequate path [...]
\end{quoting}
As we stated previously (see section~\ref{sec:uncertainty}), the Anthropocene is characterized by
major uncertainties and considerable fluctuations in climatic events. Taking these fluctuations into
account calls for the development of digital infrastructures that are frugal, robust and
adaptable. In his very stimulating proposal, O. Hamant, a biologist using biology as a case study
and as a metaphor, reminds us that the robustness of living organisms is the result of a set of
properties: \emph{heterogeneity}, \emph{randomness}, \emph{slowness}, \emph{delays},
\emph{redundancies}, \emph{inconsistencies}, \emph{errors}, \emph{incompleteness},
\emph{sub-optimality}. Following on from these properties, it seems reasonable for us to add two
important additional properties, namely \emph{locality/interactions} and \emph{emergence},
recognized by O. Hamant in~\cite{grumbach_how_2020} when dealing with complex systems.

These properties are highly desirable for the design of a computing infrastructure that should be
resistant to significant climatic fluctuations.

\section{\life Project: How Not To Worsen the Situation}

To summarize our journey, so far:
\begin{itemize}
 \item section~\ref{sec:uncertainty} pointed out several factors: the low level of available CO2 if
       we are to comply with the Paris agreement and the planetary boundaries already crossed,
       resulting in significant fluctuations in climatic events;
 \item section~\ref{sec:ict-no-limit} recalled several elements: the construction of the computer
       science disciplinary field as being limitless (even though reality massively contradicts this
       hypothesis), constantly increasing use of resources incompatible with the climate
       emergency, and  recycling that will not address the depletion of natural resources;
 \item section~\ref{sec:negcommons} proposed studying zombie technologies from the perspective of
       negative commons, which we inherit, to which we are attached and which attach us and which
       must be taken into account to ensure a fair landing of societies; in order to accommodate
       significant fluctuations, this should favor \emph{robustness \emph{and} adaptability} (and
       not \emph{adaptation} which is a static properties).
\end{itemize} 
We could summarize all these points by the desire to dispose of a digital infrastructure that is
compatible with the planetary boundaries and produces the least possible environmental damage, and
thus enables (the least possible) low-carbon computing. This is the purpose of the
\life\footnote{\life means in french ``\emph{Longévité Informatique et Frugalité Écologique}'' which
translates to ``\emph{ICT Longevity and Ecological Frugality}''.} project.

\subsection{Reintroducing Limits} \label{sec:limitsback}

As we have seen, the unlimited nature of digital technology has led to its very significant
expansion. The reintroduction of limits is therefore necessary if we want to align the use of
digital resources with planetary boundaries. It raises numerous questions:
\begin{itemize}
 \item \emph{time}: could data circulation be constrained by external resources (disposal of energy for
       example)?
 \item \emph{space}: could communication and computation be constrained by local interactions (local mesh
       network, hierarchy of data access...)?
 \item \emph{discontinuity}: could the operations only function on an intermittent basis?
 \item \emph{computation}: should everything that can be computed be computed? Could computed data be
       pre-processed and stored to reduce further identical computations?
 \item \emph{resource}: how to think in terms of (finite) stock and not (infinite) flux?
 \item \emph{availability}: should a service be always available?
 \item \emph{acceleration}: how to slow down exchanges?
 \item \emph{exhaustivity}: how to un-digitize (willingly or by necessity)?
 \item \emph{politics}: how to find the rhythm of the democratic deliberation necessary for the
       development of collective processes when digital tools progress at its own (fast)
       pace\footnote{see H. Rosa's book~\cite{rosa_acceleration_2013} for a theory of \emph{social
       acceleration}, which embraces the acceleration of technology, social change and pace of
       life; for democratic deliberation see~\cite{frenkiel_student_2024}.}?
 \item \emph{side-effects}: and how can we ensure that no rebound effect (direct or indirect) occurs?
\end{itemize}
The questions of \emph{discontinuity}, \emph{computation}, \emph{resource}, \emph{availability} and
\emph{side-effects} will be specifically considered and taken into account in
section~\ref{sec:leprojet} and following, while the others will be considered implicitly.

Reintroducing limits requires thinking in terms of non-extensible (but rather
shrinkable~\cite{maraninchi_planetary_2024,maraninchi_playing_2021}) systems and considering the
Cartesian product of
\[
\begin{array}{c}
\{ \textrm{intermittent, quotas, supply} \}\\
\times\\ 
 \{ \textrm{energy, communication, memory, computation} \} 
\end{array}
\] 
These points should be taken into account in any project that aims to consider the operating of a
digital system within the context of finite resources.

\section{The Phases of the Project} \label{sec:leprojet}

Having all those previously discussed elements in mind and in order to take into account the
constraints we discussed in the previous sections, we designed the \life project. It consists
roughly of 5 steps:
\begin{enumerate}
 \item The first step will involve collecting the widest variety of WEEE within the university,
       enabling the construction of a \ssdc (SSDC);

 \item The construction of a theoretical model to formalize the asynchronous functioning of
       heterogeneous constrained resources (in terms of computing power, communication capacity, available
       storage volume, operating systems and availability);

 \item This model will then be instantiated in the building of an effective \ssdc that unifies
       heterogeneity and allows for the implementation of a variety of information systems;

 \item Once the data-center is made available, it will be extensively tested on concrete cases to
       validate the approach taken;

 \item During all these steps, it will be important to produce relevant indicators to measure the
       direct and indirect environmental gains achieved;
\end{enumerate}
The five steps are detailed further in the following sections.

\renewcommand{\ssdc}{SSDC\xspace}
\renewcommand{\SSDC}{\ssdc}

\subsection{Waste from Electrical and Electronic Equipment as Basic Building Blocks } \label{sec:buildingblocks}

We aim at building a \ssdc{}-type infrastructure using only equipment that is considered to have
reached end-of-life (EOL). As the project is taking place within the context of the authors'
university, this will be the source all computing devices used.

\subsubsection{University as Urban Mine}

An urban mine is ``\emph{all the activities and processes involved in recovering the components,
energy and elements from the products, buildings and waste generated by human activity in the urban
environment}''~\cite{arora_potential_2017}. In our context, we restrict ourselves to WEEE. Its
purpose is to extract valuable elements through complex and energy-intensive recycling processes. In
our case, we do not consider the extraction of rare-earth elements from WEEE as it would not make
sense in our approach to minimizing environmental impact (including energy consumption). We are
ahead of the circular economy. Besides, we are at a higher level of abstraction, that of the
functionalities offered by products at the EOL but which are still fully functional\footnote{One
might wonder why these systems are being discarded. As far as our university is concerned, this is
the case when maintenance contracts expire, the equipment is no longer under warranty and the
continuity of service requires the IT Department to renew it. There is also a fad or a display
effect which, in some places, wants to renew equipment so that it is always up to date.}. Our
university, but more generally administrations and firms in the technology sector that still have
on-premise resources, are abundant \emph{mines of high-level functionalities}.

\subsubsection{Subtracting from the Technosphere}

It would have been very tempting to take the classic approach of innovation by \emph{adding} new
computing infrastructures (and new software developments). This dynamic of additive innovation could
be a classic bias in science and technology even though, quoting~\cite{adams_people_2021} (\mybold)
\begin{quoting}
Defaulting to searches for \textbf{additive changes} may be one reason that people struggle to mitigate
overburdened schedules, institutional red tape and \textbf{damaging effects on the planet}.
\end{quoting} 
In the remaining paragraphs of this article, we propose to consider an additional constraint, which
is to \emph{innovate by subtraction}~\cite{goulet_innovation_2012}\footnote{Consistently, this
reference points us back to the concepts of \emph{attachment} and \emph{detachment}, previously
discussed in section~\ref{sec:attachements} above.} at both functional and hardware level. As we
have seen, ICTs are dead technologies whose end of life will add to the large volumes of zombie
equipment. It appears crucial to maintain existing infrastructures (and thus subtracting the arrival
of new hardware) in the technosphere for as long as possible in order to delay their arrival in the
biosphere. This conservation will also have the potential benefit of not producing new
infrastructures and thus reducing their environmental impact.

In addition to this first subtraction, we propose a second subtraction in the form of the operation
of the infrastructure under conditions of severe resource constraints. This point will be addressed
later in the section~\ref{sec:SIRIUS}.


\subsubsection{Collecting the Parts}

The large quantity of equipment available (the upgrade to Windows 11 alone will result in the
disposal\footnote{One could argue that there is a very simple technical solution to switch the
entire system to this new version of Windows: install Linux and an hypervisor, then install Windows
in this environment. But for organizational reasons of the IT Department, this has not been
considered.} of at least 150 workstations - roughly 10\% of the managed fleet\footnote{Given the
size of the university's computer fleet (which has around 45,000 students), one might be surprised
at the low number of workstation changes caused by the switch to the latest version of Windows. In
fact, the 10\% in question only concerns machines that are directly managed by the central IT
Department, and the reason for the low percentage is that the policy for renewing the stock is such
that it is extremely up to date, with machines that will accept Windows 11. If the management of the
stock involved maintaining older machines, the disposal rate would be much higher.}), whether in
terms of active equipment (storage, switches, routers, WiFi access-points, backup robots, etc.) or
computing devices (desktop, laptop, servers, all-in-one devices, smartphones, single-board
computers, etc.) makes the university become an \emph{urban mine}.

Ironically, the main problem raised by this approach is not the collection of WEEE but rather the
difficulty in dealing with the phenomenal quantity of equipment available and its temporary storage
waiting to be used in the project (as well as storage for spare parts and maintenance
supplies). Both the volume of resources available and the windfall effect that the use, not the
disposal, of WEEE represents for the IT Department could very easily cause the project to expand
from a small-scale to a large-scale project, in a classic rebound effect. It should be noted that,
even though this would deplete the mine at use, the rebound effect could be eventually avoided if
the small-scale were to replace the university's official large-scale infrastructure. But we are
still a long way from that scenario!


\subsection{A Theoretical Model of Adaptable System} \label{sec:autonomous}

\begin{figure}[!h]
  \centering
  \includegraphics[width=.8\linewidth]{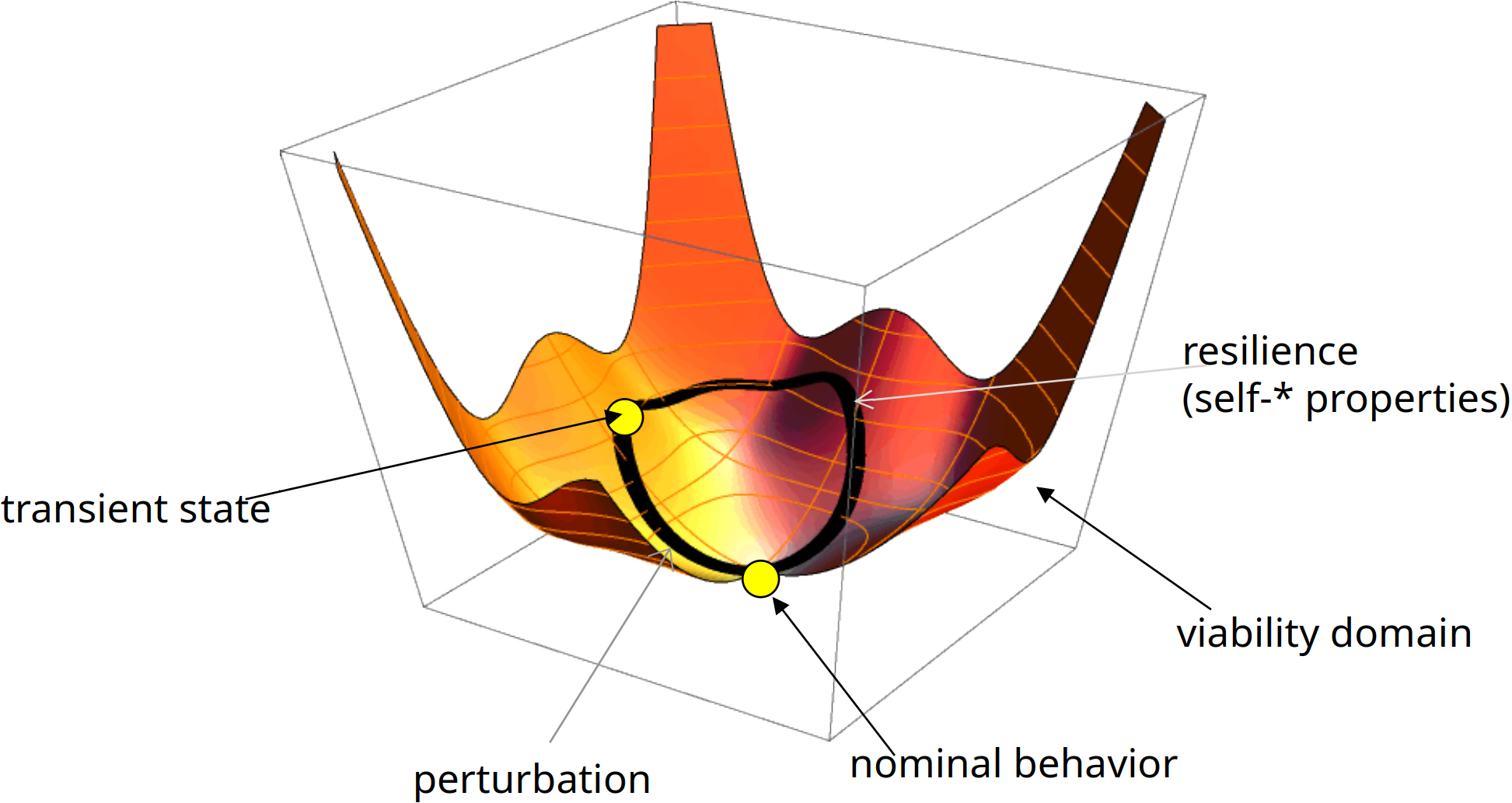}
  \caption{The small-scale data center seen as a self-adaptative dynamical system in 4 different
  states: its nominal behaviour when the system works as espected with all available resources; a
  perturbation coming from a change in the system, driven from an internal or an external
  constraint; the perturbation leads to a transient state and the self-adaption of the system to a
  new equilibirum within its viability domain.}
 \label{fig:ds}
\end{figure}

The aim is to build a system that can be adapted to a wide variety of situations resulting from
constraints external and internal to the system, according to the discussion in
section~\ref{sec:uncertainty} and~\ref{sec:hamant}: it is a dynamical system characterized by a
state, which itself is the aggregation of the states of its elements. The state evolves according to
the perturbation caused by variations of the constraints, reaching a new stable state within its
viability domain. Figure~\ref{fig:ds} describes a situation of perturbation and return to
equilibrium of the dynamical system.

Internal constraints are:
\begin{itemize}
 \item computation availability of \emph{kind} (CPU, GPU) from an \emph{element} (desktop, laptop,
       server, smartphone running Android or iOS) for a certain type of \emph{operation}
       (communication, storage, computing... resources),

 \item operational availability of hardware itself (elements may experiment failures, from the power
       supply to electronic elements on the motherboard, NIC..., leading to the unavailability of
       the element for computing purposes).
\end{itemize}
while external constraints are:
\begin{itemize}
 \item the energy available to enable the system to operate at various level (fully or partly),
 \item the addition (when a new element is avaible for the cluster) and removal (when a broken
       element has to be discarded from the cluster) of hardware elements to the data center.
\end{itemize}
To accomodate with our view of the data center as a dynamical system, we choose to follow the path
lead out by IBM in 2001 with its \emph{Autonomic Computing}
manifesto~\cite{horn_autonomic_2001,kephart_vision_2003,sterritt_concise_2005} designed to address
the software complexity crisis produced by applications that had become so large that it was no
longer possible to control the systems on which they ran. IBM's proposal was to build systems that
could manage themselves with given high-level objectives from administrators. A set of self-*
properties were defined (self-optimization, self-configuration, self-diagnosis, self-healing,
self-protecting...) ensuring the system's operational stability.

\begin{figure}[!h]
  \centering
  \includegraphics[width=.8\linewidth]{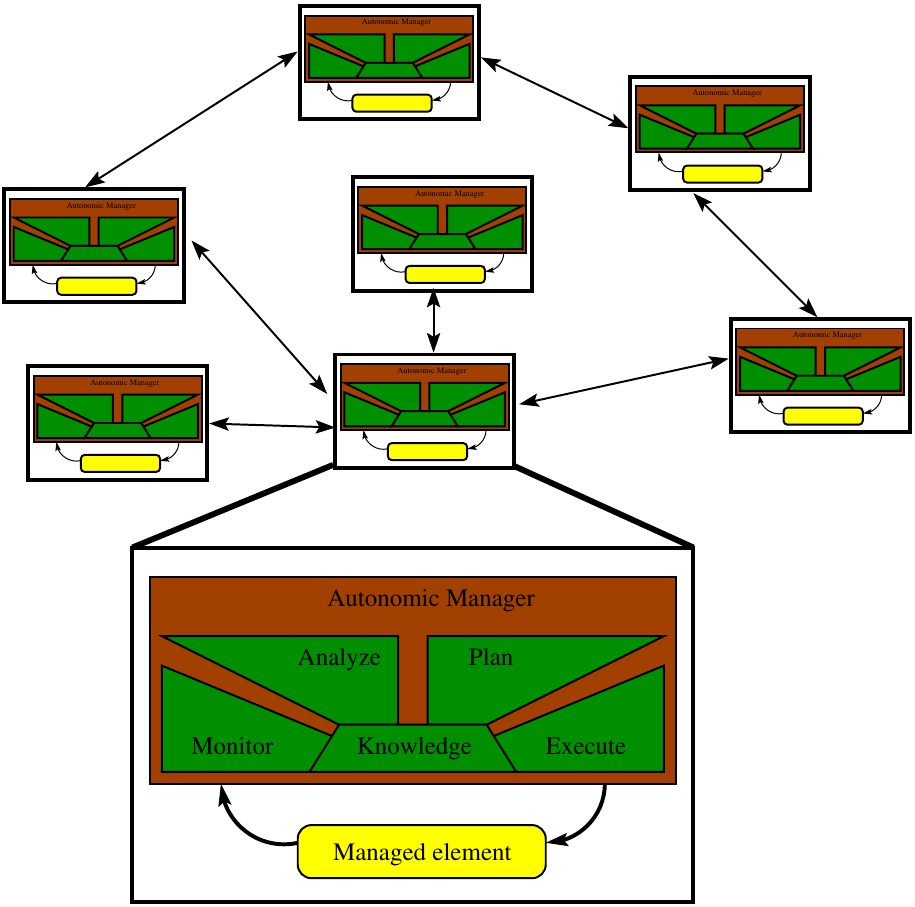}
  \caption{The structure of an autonomic element (below) where each element interacts with other elements and
with global guidelines via their autonomic managers (above). Each element has sensors and actuators allowing
to make decisions based on its current state and the global state of the system. Picture redrawn from~\cite{kephart_vision_2003}.}
 \label{fig:structure}
\end{figure}

The design of these properties will be based on a formalism of MAPE-K loops where a feedback loop
captures the state of the system and causes it to evolve according to rules. This loop consists of
``[...] sub-components for Analysis of Monitored data, Planning response actions, Execution of these
actions, all of them based on a Knowledge representation of the system under
administration''~\cite{de_lemos_feedback_2017}. These tasks are performed using informations
gathered from \emph{sensors} and produce a change in the system through \emph{actuators}, see
figure~\ref{fig:structure}.

\subsection{Putting the Pieces Together} \label{sec:together}

The \ssdc we want to build must fulfill several objectives, as described above: provide useful and
robust services to the IT Department, operate on unreliable heterogeneous hardware, and continuously
adapt to external and internal resource constraints. To achieve these objectives and in a consistent
manner with our approach, we have chosen to use only free and open source software (FOSS), tools
that are straightforward to deploy and have proven their reliability. At the software level, the
system should not require maintenance and operate as autonomously as possible, according to
section~\ref{sec:autonomous}.


For consistency, all computing elements -- desktop, laptop, servers, smartphones, single-board... --
will be referred to as \emph{nodes}.  If certain features are only available on certain devices
(like a battery for smartphone), it will be specified each time.

\subsubsection{Building an Adaptable Software Architecture} \label{sec:services}

In order to test the adaptability and robustness of the architecture, we want to provide services of
different types. To start with, we have identified usual services provided by the IT Deparment: a
Mail Transfer Agent (MTA), a cross-platform file-hosting software system, a wiki farm, simple AI
services\footnote{We do not wish to promote the development or even the use of AI, but we believe it
is important to show that meaningful projects can be carried out in this field using older
architectures and thereby show a path to cease the race for power in learning models. This sacrifice
to current trends does not, of course, blind us to the environmental issues associated with this
\emph{large technical system}.} (some of the collected devices have GPUs\footnote{Lonovo's
ThinkStation S20 have
\href{https://www.nvidia.com/docs/IO/68266/NV_DS_QFX_580_US_Mar09_FINAL_LoRes.pdf}{Nvidia Quadro FX
580 cards}, from 2009, with 512 MB of GPU memory. Its limited performance relative to current GPUs
will provide an opportunity to test the limits of our approach.} and the cluster will be used to
conduct experiments in frugal machine learning as promoted at JRAF's
conferences~\cite{trystram_journees_2025}) and high performance computing.

\paragraph{Smartphones} \label{sec:smartphones}

Smartphones are not a homogeneous class of computing devices. Within the Android-based smartphone
category, there is a wide variety of architectures and ways to access the ROM: not all devices can
be easily flashed with a new operating system. Here, heterogeneity is a source of significant
problems for simply having a functional set of computing resources. 

We envision three possibilities:
\begin{enumerate}
 \item smartphones on which a version of Linux like \emph{postmarketOS} or \emph{Ubuntu Touch} can
       be installed should be able to run the services as is\footnote{The
       \href{https://opendddb.org/en}{Database of Digital Device} lists a large number of smartphones
       and their alternative supported ROMs.};
 \item on those devices another option is by following~\cite{switzer_junkyard_2023} where
       micro-services are implemenented,
 \item and for devices where a Linux clone cannot be installed, a domain-specific framework similar
       to the \emph{Berkeley Open Infrastructure for Network
       Computing}~\cite{ries_berkeley_2012,black_exploring_2009} could be considered where tasks are
       sent to devices and results are aggregated in a backend.
\end{enumerate}
These different systems will be able to provide facilities enabling the implementation of general
services or specialized services such as data processing (as it was the case for the BOINC
framework).

Smartphones have two other distinctive features: they have GPUs and batteries. The use of GPUs will
enable testing of the suitability of simple AI projects, particularly in relation to energy
efficiency since~\cite{kwang-ting_cheng_using_2011} showed that the power consumption of a mobile
GPU is typically 2 orders of magnitude lower than its desktop version; battery management shall
include control of charge and discharge cycles based on the energy available to the system. The
presence of a battery will be a key factor to consider for the system's energy autonomy.

\paragraph{Regular Computing Devices}

When it comes to nodes as regular computers (everything but smartphones), things are much simpler
because there are many solutions available. We plan on using common and battle-tested tools:
\begin{enumerate}
 \item a Debian\footnote{\url{https://www.debian.org/index.fr.html}} distribution as Linux
       operating system;
 \item initial configuration using PXE\footnote{\url{https://wiki.debian.org/PXEBootInstall}} on
       the NIC together with DHCP and a TFTP server, SSH server and
       Ansible\footnote{\url{https://docs.ansible.com/}};
 \item virtualization using native Docker\footnote{\url{https://www.docker.com/}},
       ProxmoxVE\footnote{\url{https://www.proxmox.com/en/}},
       Kubernetes\footnote{\url{https://kubernetes.io/}};
 \item backup using Proxmox Backup
       Server\footnote{\url{https://www.proxmox.com/en/products/proxmox-backup-server/overview}} and
       Restic\footnote{\url{https://restic.net/}} on RAID devices and a collected tape
       robot\footnote{An Overland Storage NEO 2000e from 2013, see
       \url{https://support.bull.com/ols/product/storage/lib/overland/lxn-neo-e/neoe2k4k}.};
 \item monitoring with Prometheus\footnote{\url{https://prometheus.io/}} and
       Grafana\footnote{\url{https://grafana.com/grafana/}}.
\end{enumerate}

\paragraph{Storage}

A key element of any distributed computing project that requires a minimal level of robustness is
data storage. To achieve this, we will rely on the infrastructure provided by Garage\footnote{A
presentation is available at
\href{https://archive.fosdem.org/2024/schedule/event/fosdem-2024-3009-advances-in-garage-the-low-tech-storage-platform-for-geo-distributed-clusters/}{FOSDEM's
2025 edition}.}~\cite{deuxfleurs_association_garage_2025} which provides all the desired features in
an unreliable environment: ``it's a lightweight geo-distributed data store that implements the
Amazon S3 object storage protocol. It enables applications to store large blobs [...]  in a
redundant multi-node setting.''. It is highly resilient to network failures, network latency, disk
failures, and sysadmin failures -- all very likely events in our project.

\paragraph{Networking}

To ensure maximum network robustness, redundancy will be implemented using a pair of refurbished 48
ports Avaya Nortel 4548GT-PWR switches\footnote{The university has a very large number of these
switches, which are currently being replaced with newer versions following a change of the backbone
network. All models are yet fully functional, the previous backbone was not congested with low
latencies and only a low percentage of used bandwidth, but various local socio-technical constraints
led to this choice.}. The nodes will be connected differently depending on their type:
\begin{itemize}
 \item Smartphones: they have two (actually three, but we do not consider cellular here) distinct
       means of communication, wireless and Ethernet connection via USB port. While wireless could
       be a fallback solution in case of network failure, a USB port connection is preferable
       as~\cite{na_scalable_2021} have shown significant limits to the use of wireless.

 \item Non-smartphones: depending on the number of network cards on each node, they will be
       connected to one or both switches to ensure redundancy in the event of a malfunction (of the
       switch or node). Nodes with only one network card will be distributed across each switch. The
       failover from one subnet to another will be automatic.
\end{itemize}
A degraded operating scenario in which switches no longer function will take into account the
construction of a point-to-point mesh network~\cite{iera_making_2011}. This infrastructure will
enable the implementation of the \emph{amorphous computing} model discussed below.

\subsubsection{Coupling Heterogeneous Systems for Robustness and Autonomic Computing}

While the use of smartphones (mainly Android-based) or used computers for cluster computing is
nothing new~\cite{busching_droidcluster_2012}, their heterogeneity (all studies mentioned so far
only deal with one smartphone model, mainly Fairphones) and simultaneous use, taking into account
adaptability and maximized autonomy, presents a number of challenges. The dynamic addition and
removal of resources is one of the main constraints. Since the system must enforce self-*
properties, it will be necessary to
\begin{enumerate}
 \item identify the capabilities (computing power, memory, bandwidth, etc.) available for each device;
 \item calculate each device's \emph{efficiency} in the form of a ratio \emph{computing power} by
       \emph{energy consumed} (in the form of mips and mflop per joule);
 \item map and schedule the tasks (ranging from high-level tools like an MTA to low-level data
       crunching in the form of micro-services) in order to maximize the use of computing resource
       with respect to available energy.
\end{enumerate}
Implementing the autonomic manager described in figure~\ref{fig:structure} of
section~\ref{sec:autonomous} requires local monitoring at the device level as well as
globally. Initially, centralized monitoring (with redundancy) will be implemented to closely track
the system's operation. Since the failure of a component may happen at anytime, devices will be
continuously monitored by a watchdog mechanism, which will trigger, if required, a rebalancing of
available resources. In a second phase, a decentralized task orchestration mechanism will be
implemented to meet the autonomy and adaptability requirements. Moreover, a large amount of work has
been done on self-repair strategies for autonomous systems based on FPGAs that could be of
interest~\cite{salvador_fault_2011}.

A direction we would like to explore is related to the \emph{Amorphous Computing} (AC)
project~\cite{abelson_amorphous_2000}. AC aimed at solving problems using a large set of simple and
unreliable computing devices with very limited communication capabilities and only local
interactions. These devices could break down at any moment, but this should not affect the current
computation. While many low-level algorithms have been defined using the principles of AC we believe
that the studies in that field should help us design, at a higher level of abstraction, the behaviour
of the \ssdc in a severely degraded mode where the only possible communications are peer-to-peer and
only low-level functions can be performed.

%
%

\subsubsection{Ensuring Optimal Functionality Under Energy Constraints} \label{sec:optienergie}

The energy required to operate the system (the boundary of the system is considered to be the
connection to the network of the entity hosting the data center, here our university) will
potentially come from multiple sources: the electrical grid, uninterruptible power supply, and
renewable energy sources such as photovoltaic panels (we already have collected
\emph{uninterruptible power supply} (UPS) and only second-hand \emph{photovoltaic panels} (PV) will
be considered). When \emph{variable renewable energy} is considered, two types of problems arise:
how to manage variations in energy availability \emph{at the node level} and \emph{at the whole
system} level.

\paragraph{Variable Renewable Energy at the Node Level.} \label{sec:SIRIUS}

At the node level, the intermittent nature of energy availability has to be taken into account, as
this is an external constraint on the system that will determine its availability. There is a
significant literature on this topic, which is based primarily on two approaches: at the node level,
a decomposition of the application into tasks, after which a backup of the system state is made, or
the insertion of checkpoints into the
application~\cite{umesh_survey_2021,bernabeu_cost-optimal_2023,bernabeu_support_2023}. Both
approaches will be considered depending on the nature of the services offered to the users.

Another approach being considered is the one developed as part of the SIRIUS\footnote{That
translates into \emph{Resilient, Useful, and Sober Information System}, a project developed in an
engineering school to take into account the highly probable scarcity of resources in the near future
and the control of the rebound effect, in a nutshell, ``\emph{computing within limits}''.}  educational
project: depending on resource constraints, an application is dynamically stripped of features to
reduce its consumption. This approach requires questioning the purpose of each function served, its
importance, and the possibility of making it temporarily (or permanently) unavailable. This is a
concrete example of what was outlined in the questions on the reintroduction of limits in
section~\ref{sec:limitsback}, more precisely, an \emph{ecology of functionality}.

\paragraph{Variable Renewable Energy at the Whole System Level.}

At the system level, fine-tuned management of energy availability will require predictive analysis
based on weather forecasts in order to anticipate energy availability cycles. This anticipation,
combined with the energy profiles of nodes and tasks, will make it possible to anticipate system
availability (that is, which nodes are up) but will require selecting which functions will be
unavailable (between the MTA and the AI application that analyzes data, which one should be shut
down?\footnote{This seemingly mundane question will be at the heart of future debates when resources
become scarce and choices must be made about what to keep and what to abandon or \emph{close}, to
use A. Monnin's terminology, see section~\ref{sec:negcommons}
and~\cite{bonnet_heritage_2021}.}). This could be done using classical multi-objective optimization.

Another way to address the problem is to keep only the most sober nodes -- for this notion see
section~\ref{sec:impact} on the impacts below -- and move applications that are considered a
priority to these nodes, if they are compatible. Indeed, some nodes like smartphones cannot handle
all types of activity.

\paragraph{Variable Renewable Energy with Smartphones.}

As stated in section~\ref{sec:smartphones}, smartphones having batteries means that they have both
an additional source of energy available and that the available energy must be anticipated: in
situations of energy shortage, we can imagine that the system would operate solely on the energy
available to smartphones (at the risk of reduced performance~\cite{corentin_libert_ecole_2024}), but
then only tasks eligible on these devices could be considered, conserving the scarce energy
available for system administration (data collection, routing, orchestration...). Furthermore,
smartphone batteries could also be used as a supplement to UPSs in a similar way to the
\emph{vehicle to grid}~\cite{tan_integration_2016} technology, allowing bidirectional energy
exchange between electric vehicles and the power grid. This would be a scaling-up of what is already
possible using one smartphone as an induction charger for another device (many smartphones already
have this possibility using the Qi standard). Here, the whole system would be powered by energy
available in smartphone batteries (and we may even dare to dream that one day, institutions such as
universities will have their own smart-grid, with PV panels and distributed management of energy
production and consumption).


\subsection{Robustness} \label{sec:robustness}

The robustness of the \ssdc will be a key element of the project to validate its ability to adapt to
faulty nodes, intermittent power supply, while still providing meaningful services. As adaptability
is more important than efficiency, particular care will be taken to ensure minimum operation, even
in severely degraded conditions, as outlined in the previous sections. Whether through the use of
uninterruptible power supplies and/or smartphones, or mesh network operation in AC mode, we shall be
able to test the resilience of the system and the ability of the dynamical system to reach a stable
state and return to a new stable state after a major system disruption.

Heat generated by the \ssdc has to be taken into account. Smartphones are not a problem because they
are designed not to heat up beyond a certain threshold, which, if exceeded, causes the smartphone to
shut down. Furthermore, the heat generated is very low and a regular fan is sufficient to maintain
an operating temperature~\cite[p. 405]{switzer_junkyard_2023} for over a hundred smartphones. The
situation will be different for network equipment and servers. To this end, we are working with
facilities management and ancillary services to study how to use the heat produced to heat buildings
(in cold seasons) and how to make positive use of the heat produced in the hot seasons. Of course,
excessive temperatures will have to be taken into account in order to shut down part (or all) of the
\ssdc if necessary.

The use of WEEE will probably lead to a large number of failures\footnote{This may well not be the
case given the age of the machines collected, some of which date back to 2009 and are still in
perfect working order. Maintenance of these machines will also be planned, as breakdowns are often
due to a malfunctioning capacitor on the motherboard, which is easily replaceable.} that will need
to be evaluated to establish the relevance and sustainability of the approach. However, these
failures should be offset by the very large number of machines available, as long as the failure
rate relative to available stock remains favorable. Robustness also refers to the system's ability
to meet its commitments to reduce environmental impact, a point that is discussed in the following
section.

It will also be necessary to continue discussions with the IT Department on the relevance of working
with WEEE. Many French institutions (universities, hospitals, etc.) have been the target of hacking
attacks that have shut down their IT systems. The \ssdc could also be a solution to this type of
situation, where minimum services need to be quickly restored while the information system is being
repaired (which generally takes several months).


\subsection{Assessing the Impacts} \label{sec:impact}

To determine whether our approach to the significant reduction in the impact of a \ssdc, we need to
assess both the impact of setting up and using the data center and the impact of what has been
avoided by using it. 
We consider potential gains coming from three sources:
\begin{enumerate}
 \item equipment that was not purchased due to the extended use of existing equipment,
 \item increased efficiency through the additional use of specific devices (that is, smartphones) or the first life of the node,
 \item a reduction in the number and use of computing devices through a change in mindset that embraces digital frugality.
\end{enumerate}
We review these gains which ultimately appear artificially distinct because they are so intertwined.

\subsubsection{Unpurchased Equipment} \label{sec:unpurchased}

We want to compare the following two situations: a subset of the IT Department's digital
infrastructure that provides certain services and the \ssdc that implements these very same services
(for example those identified in section~\ref{sec:services}). First, a precise inventory of the IT
Department's digital equipment must be carried out and the target services identified. Next, 
life-cycle assessment\footnote{The reader not familiar with the concepts of LCA should consider
reading~\cite{loubet_life_2023} for a very good introduction to the matter, applied to digital
devices.} (LCA) must be carried out, for each service X, with functional unit ``\emph{provide
service X for Y users over Z months}'' on the IT Department's equipment and on the \ssdc. We
consider following the methodology of~\cite{samaye_life_2025}.

A quick analysis of the situation supports our belief that the \ssdc should have a lower
impact. While the increased efficiency of newly bought infrastructures to replace older devices is
beyond doubt (see section~\ref{sec:laws}), it appears, however, that according
to~\cite{pirson_environmental_2023}, the environmental impact (in terms of energy, carbon footprint,
and water) for integrated circuits was not significantly reduced in the 1980-2010
period\footnote{Meanwhile, the total silicon area produced grew by 3.6\% per year, leading the
authors to call for sufficiency.}. In fact, for the \ssdc, all phases prior to its use phase will be
allocated to the IT Department (the equipment was purchased and used in the first phase of its
life), and only the second use phase will be allocated to the \ssdc (after an initial use phase of
approximately 5 to 7 years). Furthermore, given the electrical mix in France where the project takes
place is very low-carbon intensive\footnote{According to
\href{https://app.electricitymaps.com/zone/FR/72h/hourly}{ElectricityMaps}, at the time of writing,
in 2025, April the $\textrm{30}^\textrm{th}$, about 28 CO2eq/kWh.}, the GHG impact should be minimal.

\subsubsection{Increased Efficiency}

For the smartphone part of the \ssdc, the authors of~\cite{switzer_junkyard_2023} define
\emph{Computational Carbon Intensity} (CCI) as a ratio of GHG production (manufacturing phase + use
phase) and computation. While this measure highly depends on the electricity mix used to power the
devices and their lifespan, a smartphone-based system is 9.8 to 18.9 times more CCI efficient, after
3 years of use, depending on the nature of the application.  Due to the first life of the node, this
measure will also be very favorable for conventional computers.

CCI is a very good metric for the project, and we propose to extend it to take into account all the
impacts revealed by a life cycle analysis -- the 17 ReCiPe 2016 Midpoint (H) categories -- and to
adapt its calculation to take into account the first life phase of the node used in the \ssdc.

\subsubsection{Digital Frugality}

An important side effect of increasing the lifespan of equipment is that it frees up financial
resources that can be reallocated to other areas, such as social programs, improving student life,
building accomodations, insulating buildings... and possibly closing (or re-shaping) courses
considered to be negative commons. One could even imagine a positive feedback loop that would lead
university teaching and services to use fewer IT resources, thereby reducing the use of the \ssdc
itself, paving the way toward frugality.

\section{Related Work and Concerns}

\subsection{Related Work}

\subsubsection{On a Technical Side}

There is a large number of works, some of them old and others more recent, on related issues. Among
these, the interested reader may wish to look at the following works: when it comes to 
computing with the least possible energy~\cite{de_valk_permacomputing_2022};
grid computing on mobile devices~\cite{goos_ourgrid_2003,vishalan_grid_2006,black_exploring_2009,hosseini_crowdcloud_2017,yasrab_grid_2019};
intermittently available energy~\cite{raghavan_intermittent_2012,liu_unpacking_2023};
using virtual machines to overcome closed architectures~\cite{black_exploring_2009};
mobiles devices as computing devices and edge-computing~\cite{shahrad_towards_2017,hirsch_augmenting_2018,switzer_junkyard_2023,corentin_libert_ecole_2024,perseverance_ngoy_supporting_2025}
or using microservices to access resources~\cite{gan_open-source_2019};
AI on smartphones~\cite{libert_low-cost_2024} using Kubernetes and TensorFlow, using OpenMPI~\cite{na_scalable_2021}.

Several recent studies are moving in a similar direction to ours, at a smaller scale. The work
of~\cite{sutherland_strategies_2022} explores strategies for degrowth and energy autarky using solar
panels and supercapacitors to store energy for computing devices. The work
by~\cite{switzer_junkyard_2023}, already mentioned several times focuses exclusively on identical
smartphones of the same model-year. We included smartphones in our approach (because we aim at
dealing with heterogeneity and there is a massive amount of home-stored but unused smartphones), but
due to the specific nature of our urban mine, not only do we have a lot of diversity among the
collected devices, but they are also not the main resource at our disposal. However, as part of the
\href{https://ewwr.eu/}{European Week for Waste Reduction}, we
\href{https://slow-tech.fr/life#la_collecte_de_deee}{organized a campaign to collect used
smartphones}.

\subsubsection{Broadening the Discussion}

The approach proposed by~\cite{samaye_towards_2024,da_silva_optimization_2023} is particularly
interesting since it combines equipment whose lifespan has been extended and uses photovoltaic
panels as source of energy in an edge-computing model. There is an added twist: hosted virtual
machines migrate according to the availability of solar energy, which itself circulates between data
centers. This combined approach could be adapted to a university's campuses such as ours, which
are more than 80-km distant from North to South (at least in terms of VM traffic, as solar energy is
local)\footnote{But in a way, it breaks down our wish to bring back limits in time/space and
discontinuity in~\ref{sec:limitsback}. This should be investigated.}.

In an academic setting, the approach of~\cite{angeli_conceptualising_2022} aims, through the
RECLUSTER project, to re-internalize their cloud services, regain control of the infrastructure, use
FOSS, and reduce WEEE by reusing equipment that was left on shelves and no longer in use. This means
that control over tools is regained, less resource-intensive software can be used, and big tech
services can be set aside. Ultimately, self-hosting allows to return to a situation similar to what
universities were like before the outsourcing movement. We did not include their concern about the
desire to reinternalize the services that universities have outsourced because this is not where our
project originates. It seemed to us that the reasons given in section~\ref{sec:uncertainty} and more
specifically in section~\ref{sec:ictgrowth} call for  drastic change in the way we view and use
computing architectures and WEEE. However, we fully agree with the authors' proposal and endorse
their approach.


\subsection{Related Concerns}

What sets the \life project apart from all these approaches is its origin, which stems from  deep
frustration at being part of the problem (both in terms of the transmission of IT knowledge and the
use of computing resources). Without looking the other way, we felt it was necessary to come up with
a way of maintaining services with the lowest possible impact, which could eventually be reduced to
zero, if needed. Technically, we propose using heterogeneous, unreliable hardware\footnote{We
clearly are in the category of \emph{hardware sufficiency} as defined
in~\cite{santarius_digital_2023}.} (including smartphones, computers, and active equipment) and
unreliable storage media, which allows us to provide services intermittently, potentially reducing
functionality, while remaining adaptable and composed entirely of WEEE.

\subsubsection{Computing Meets Low Tech}

The low-tech community considers that placing ICT and low tech in the same sentence is an
oxymoron~\cite{bihouix_lage_2014,mateus_perspectives_2023}. Without getting into the controversy
over whether digital technologies can(not) be low-tech, it can be seen that the properties that we
want in our system share common points with those of low-tech. Low tech~\cite{ademe_demarche_2022}
advocate for (we only keep the properties what our project shares with low tech):
\begin{itemize}
 \item \emph{sustainability} with low environmental impact and reduced consumption on resources;
 \item \emph{autonomization} by reducing interdependencies;
 \item \emph{locality} by reducing resource pressure;
 \item \emph{accessibility} by favoring robustness, cost-efficient system with increased longevity.
\end{itemize}
It is comforting to see that the project we are proposing is in line with the low-tech movement,
which places frugality and sustainability at the heart of its approach.

\subsubsection{Inverse Legacy Problem}

Among the equipment collected, we have a few \emph{Lenovo ThinkStation S20 model 4157}
servers\footnote{\url{https://pcsupport.lenovo.com/ec/fr/products/workstations/thinkstation-s-series-workstations/thinkstation-s20}}. These
servers are from 2009 and have Intel Xeon X3503 Microprocessors CPUs with Nehalem architecture. This
architecture predates the Sandy Bridge architecture~\cite{rotem_power_2011} released in 2011, and
does not have RAPL instructions for estimating CPU
consumption~\cite{noureddine_analyzing_2024}. Although fully functional, these computers will not be
able to execute code involving this instruction set. The situation is similar for the operating
system, where there is no guarantee that Linux (in our case) will continue to include drivers for
(very) old machines.

This situation raises the opposite problem of \emph{legacy systems} that need to run older software
on newer architectures (including CPUs, operating systems, and libraries). Here, the goal is to
ensure that \emph{newer code can continue to run on older hardware} -- it is an essential requirement
of the project. To our knowledge, this issue has never been addressed before even though it will
become increasingly crucial in the coming decades. We propose to call this situation the
\emph{inverse legacy problem}\footnote{This notion differs from \emph{regular obsolescence} in that
it is a forward looking movement of unsupported technology, whereas inverse legacy is a
backward-facing movement of unsupported technology.}.

\subsection{Conclusion}

This work has been guided by the twofold concern of an increasingly dramatic environmental situation
and, at the same time, the need to provide computing resources for a society strongly attached to
digital technology. We established that digital technology's current contribution to aggravating
environmental issues was not compatible with the Paris agreement, and that recycling could not
address the constraints linked to the reduced availability of resources. To find a
suitable path through this set of constraints, we called upon the concepts of \emph{negative
commons}, \emph{zombie technology}, \emph{attachments/de-attachments/re-attachments}, and advocated
the emergence of \emph{adaptable} and \emph{limited digital technologies}.

We have sketched out a project that will \emph{never become} \emph{old-fashioned}, because it is already
old-fashioned \emph{by design}, based on building blocks made entirely of WEEE from our urban mine
at hand. The small data-center under construction, modeled as an autonomous dynamic system, will be
powered by renewable energy. A study of the reductions achieved in terms of the system's
environmental impact should validate the approach.

We'd like to conclude with a plea for research on digital technologies that takes the notion of
planetary boundaries very seriously, avoiding their aggravation; that explores neglected research
trajectories; that prioritizes robustness and adaptability over performance; that changes our
perspective on waste; that questions our imaginations; and that places digital technologies at the
center of the political debate. Only then will we be able to begin the process of digital degrowth,
which we believe to be necessary.



\section{Acknowledgments}

This work has been partially funded by the
ERASME 
project and the ``\emph{Chaire délibération}'' at Université Paris-Est Créteil, ANR France 2030.

The authors wish to thank 
A. Brenner, 
E. De Carvalho, 
A. Fantou, 
G. Giraud. 
G. Hains, 
L. Joly, 
H. El Karmouni, 
C. Marquet,
C. Papin,
E. Pelz, 
S. Quinton,
G. Samain,
L. Souidi, 
M. Souret, 
E. Trublereau,
P. Valarcher, 
the members of the GDRS ÉcoInfo,
the students at EPISEN's SI Dept,
the students at Fontainebleau's IEP
for fruitful discussions, their open-mindedness and sharp wit. We would also particularly like to
thank the reviewers of this work, who made a large number of very valuable suggestions for improvement.

\bibliographystyle{ACM-Reference-Format-num} 
\bibliography{2025-06_LIMITS}

\end{document}